\documentclass[aps,prl,twocolumn,superscriptaddress,showpacs]{revtex4}

\usepackage{graphicx}
\usepackage{endnotes}
\usepackage{color}
\usepackage[FIGTOPCAP, hang, tight ]{subfigure}
\usepackage{amssymb}
\usepackage{setspace}

\begin{document}

\title{Electronic Raman Scattering in Metallic Carbon Nanotubes }

\author{H. Farhat}
\affiliation{Research Laboratory of Electronics, MIT, Cambridge MA 02139}
\author{S. Berciaud}
\email{stephane.berciaud@ipcms.unistra.fr}
\affiliation{Departments of Physics and Electrical Engineering, Columbia University, New York, NY, 10027}
\affiliation{IPCMS (UMR 7504), Universit\'e de Strasbourg and CNRS, F-67034 Strasbourg, France}
\author{M. Kalbac}
\affiliation{J. Heyrovsky Institute of Physical Chemistry, Academy of Sciences of the Czech
Republic, V.V.i., Dolejskova 3, CZ-18223 Prague 8, Czech Republic,}
\author{R. Saito}
\affiliation{Department of Physics, Tohoku University, Sendai, 980-8578, Japan}
\author{T.F. Heinz}
\affiliation{Departments of Physics and Electrical Engineering, Columbia University, New York, NY, 10027}
\author{M.S. Dresselhaus}
\affiliation{Department of Electrical Engineering and Computer Science, MIT, Cambridge MA 02139}
\affiliation{Department of Physics, MIT, Cambridge MA 02139}
\author{J. Kong}
\email{jingkong@mit.edu}
\affiliation{Department of Electrical Engineering and Computer Science, MIT, Cambridge MA 02139}
\date{\today}


\begin{abstract}
We present experimental measurements of the electronic contribution to the Raman spectra of individual metallic single-walled carbon nanotubes ($M{\rm -SWNTs}$).  Photo-excited carriers are inelastically scattered by a continuum of low-energy electron-hole pairs created across the ``graphene-like'' linear electronic subbands of the $M{\rm -SWNTs}$.  The optical resonances in $M{\rm -SWNTs}$  give rise to well-defined electronic Raman peaks. This \textit{resonant} electronic Raman scattering is a unique feature of the electronic structure of these one-dimensional quasi-metals.

\end{abstract}

\maketitle

Inelastic light (Raman) scattering is a versatile tool for studying elementary excitations (phonons, electron-hole pairs, plasmons, magnons) in condensed matter systems. In particular, \textit{electronic} Raman scattering  has provided insight into electronic behavior in low-dimensional semiconductors \,\cite{pinczukLSS1989} and strongly correlated materials \,\cite{devereauxRMP2007}.  In recent years, metallic single-walled carbon nanotubes ($M{\rm -SWNTs}$) have emerged as a unique medium for studying electrons and their correlations in one dimensional systems \,\cite{deshpandeNATURE2010,deshpandeSCIENCE2009,wangPRL2007}.  However, despite the large impact of vibrational Raman spectroscopy in the area of carbon nanotube research \,\cite{dresselhausPhysRep2005}, electronic Raman scattering has not yet been observed in these structures. 
The Raman features that have been studied to date in carbon nanotubes are associated with the phonon modes that can scatter light due to electron-phonon coupling~\cite{dresselhausPhysRep2005}. The optical resonances $S_{ii}$ and $M_{ii}$ in semiconducting ($S{\rm -SWNTs}$) and metallic nanotubes, respectively, produce incoming (outgoing) \textit{phonon} Raman resonances, when the energy of the exciting (scattered) photon matches that of an optical transition\,\cite{dresselhausPhysRep2005}. This resonant enhancement makes it possible to observe Raman scattering down to the single nanotube level~\cite{jorioPRL2001, meyerPRL2005}.

As opposed to semiconducting species, metallic nanotubes possess a pair of linear subbands.  They arise from the well-known gapless dispersion of graphene near the corners of its Brillouin zone~\cite{saitoBOOK, TAP}. Low-energy electron-hole pairs across these bands can be formed, e.g., from the decay of a phonon (Landau damping) or from the inelastic scattering of light, via the Coulomb interaction. The former mechanism reduces the phonon lifetime and has been extensively studied~\cite{lazzeriPRB2006, farhatPRL2007,wuPRL2007}. The latter corresponds to \textit{electronic} Raman scattering (ERS) and will be discussed below.

\begin{figure}
		\includegraphics[width=0.5\textwidth]{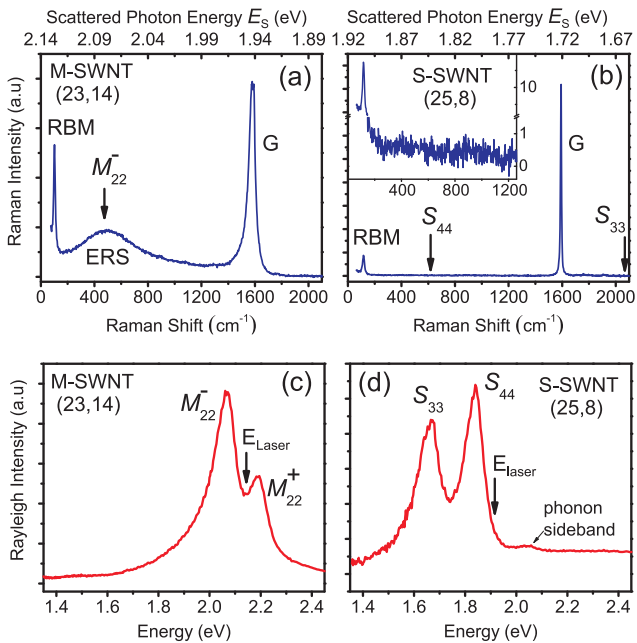}
		\caption{
(Color online) The Raman $\rm(a,b)$ and Rayleigh $\rm(c,d)$ scattering spectra are used to identify the $(n,m)$ indices of an isolated $M{\rm -SWNT}$ in (a) and (c) and an isolated $S{\rm -SWNT}$ in (b) and (d).  In (a) and (b), the Raman shifts are indicated on the lower axis and the corresponding energies of the scattered photons are indicated on the top axis.  The electronic Raman scattering (ERS) feature is the broad peak at $\sim500~{\rm cm^{-1}}$ in (a). Labels in (c) and (d) indicate the optical transitions, while the laser energy for the Raman measurements are indicated with arrows. Likewise, arrows in the Raman spectra (a,b) indicate the energies of the optical transitions ($M_{22}^{-}$, $S_{33}$, $S_{44}$) as obtained from fitting the Rayleigh scattering spectra using an excitonic model~\cite{berciaudPRB2010}.  The $M_{ii}$  transitions are split into $M_{ii}^{+}$ and $M_{ii}^{-}$ due to the trigonal warping effect~\cite{saitoPRB2000}.  The inset of (b) shows the weak and featureless low-energy tail in the Raman spectrum of the $S{\rm -SWNT}$ in (d).}
\label{Rayleigh-Raman}
\end{figure}

In this letter, we report on a new feature, found exclusively in the Raman spectrum of $M{\rm -SWNTs}$, that we attribute to ERS from low-energy electron-hole pairs.  Figure ~\ref{Rayleigh-Raman}(a) shows the Raman spectrum of a $M{\rm -SWNT}$ with the labeled ERS feature located between the G-mode and the radial breathing mode (RBM) features.  Remarkably, we observe well-defined spectral peaks rather than a flat background, as would be expected for scattering by a continuum~\cite{cerdeiraPRB1973}.  In contrast to the narrow phonon features, the relatively broad electronic peak is highly dispersive and appears at a \textit{constant scattered} photon energy in the Raman spectrum. We attribute this phenomenon to a resonant enhancement of the ERS that occurs when the energy of the scattered photon matches one of the $M_{ii}$ excitonic transition energies~\cite{TAP,wangPRL2007, wuPRL2007, berciaudPRB2010, sfeirSCIENCE2006}.  No such feature is observed in the Raman spectrum of $S{\rm -SWNTs}$ that are excited under similar conditions.  The electronic Raman feature in $M{\rm -SWNTs}$ can be quenched by shifting the Fermi energy above the energy of the electrons or holes involved in the interband transitions.  

The ERS feature, may have previously been overlooked because of the difficulty in distinguishing a spectrally broad signal from an extrinsic background from the substrate supporting the SWNTs. To overcome this problem, we have performed key experiments on isolated suspended SWNTs, which were grown by chemical vapor deposition over open trenches~\cite{supp-info}. Such SWNTs are therefore free from environmental perturbations and substrate scattering; however, we also show that ERS is observable for $M{\rm -SWNTs}$ lying on a substrate. 
  
To demonstrate that the Raman peaks which we attribute to ERS are intrinsic features of individual $M{\rm -SWNTs}$, we have focused on suspended SWNTs whose $(n,m)$ chiral indices have been identified by means of combined Rayleigh and Raman scattering spectroscopy~\cite{sfeirSCIENCE2004, sfeirSCIENCE2006, supp-info,wuPRL2007, berciaudPRB2010}. The Rayleigh scattering spectrum of a SWNT gives its optical transition energies and allows one to distinguish between an isolated SWNT and a SWNT bundle.  Along with the information obtained from the RBM and the zone-center optical phonon modes (G-mode) in the Raman spectra, the optical transition energies can be used to make a tentative assignment of the $(n,m)$ indices of a given isolated SWNT~\cite{supp-info}.  In Fig.~\ref{Rayleigh-Raman}, we compare the Raman and Rayleigh spectra of a $M{\rm -SWNT}$ $\left[\rm \left(mod(\textit{n-m}),3\right)=0\right]$ and an $S{\rm -SWNT}$ $\left[\rm \left(mod(\textit{n-m}),3\right)=1,2\right]$  whose optical transition energies and structural parameters were thus identified. 

For this particular $M{\rm -SWNT}$, the ERS feature occurs at a Raman shift of $\sim 500~{\rm cm^{-1}}$ ($\sim 62~{\rm meV}$).  The corresponding scattered photon energy of the new feature in Fig.~\ref{Rayleigh-Raman}(a)  (at $2.08~{\rm eV}$) matches the position of the $M_{22}^-$ transition (obtained from the Rayleigh scattering spectrum in Fig.~\ref{Rayleigh-Raman}(c).  This peak is symmetric and has a full width at half maximum (FWHM) of $\sim620~{\rm cm^{-1}}$($\sim 77~{\rm meV}$). For the case of $S{\rm -SWNTs}$ also excited just above an excitonic transition,  the Raman spectrum does not exhibit any feature at the energy of the same excitonic transition (Fig.~\ref{Rayleigh-Raman}(b)), nor is there an appreciable inelastic scattering background.

\begin{figure}
	\includegraphics[width=0.5\textwidth]{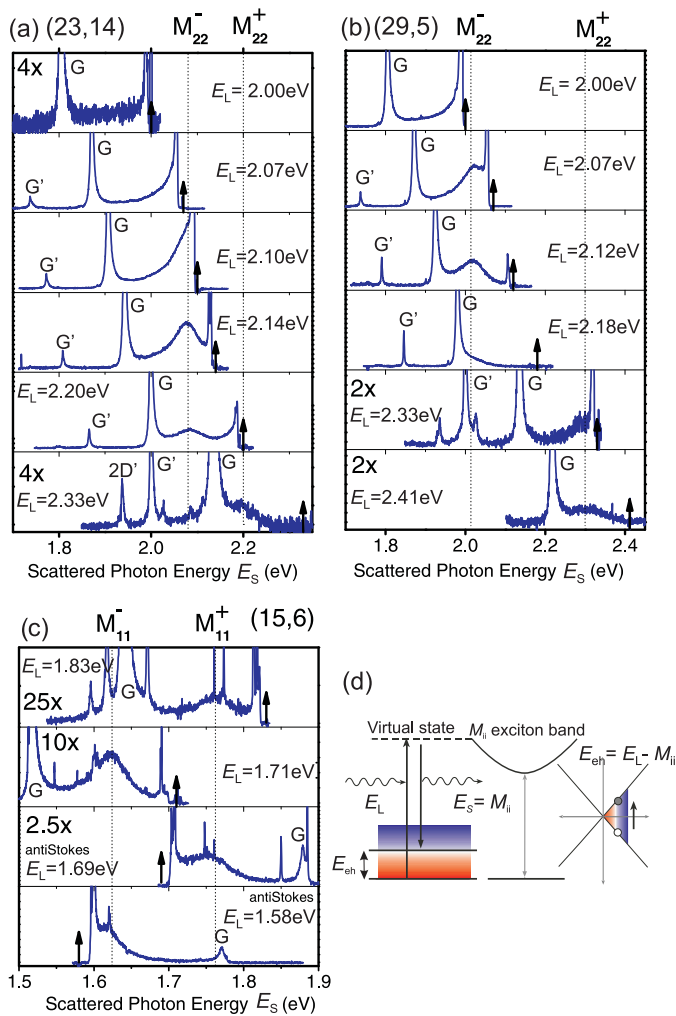}
		\caption{
(Color online) (a) Raman spectra  of the (23,14) $M{\rm -SWNT}$ from Fig.~\ref{Rayleigh-Raman}(a) and (c), plotted as a function of the \textit{scattered photon energy} $E_{\rm S}$.  Each spectrum is collected at a different laser photon energy $E_{\rm L}$, as indicated by the position of the black arrows. The ERS features are the broad peaks that line up with the dotted vertical lines, which indicate the energies of the $M_{22}$ transitions.  The positions of the G-($\sim1580~{\rm cm^{-1}}$), G'-($\sim2700~{\rm cm^{-1}}$) and 2D'-mode ($\sim3150~{\rm cm^{-1}}$)  features \cite{dresselhausPhysRep2005} are indicated on the spectra.  All spectra are normalized to the integrated G-mode intensity. (b) Same as (a) but for a (29,5) isolated $M{\rm -SWNT}$.  (c) Spectra from a $M{\rm -SWNT}$ on a substrate (not identified by Rayleigh scattering).   Here the ERS feature is shown for the $M_{11}$ peaks and appears on both the Stokes and anti-Stokes spectra. From the RBM frequency ($160 {\rm cm}^{-1}$) and the position of the ERS peaks, we assign this $M{\rm -SWNT}$ as (15,6). (d) Schematic of resonantly enhanced electronic Raman scattering in an armchair $M{\rm -SWNT}$.  A continuum of e-h excitations (illustrated by the color gradients on the left and right panels) may scatter the incident photon of energy $E_{\rm L}$. The downward arrow in the left panel indicates the resonantly enhanced scattering event in which the outgoing energy $E_{\rm S}$ matches the $M_{ii}$ optical transition energy (middle panel).  The full and empty circles in the right panel, illustrate the e-h pair that is resonantly selected from a continuum of e-h excitations.
	}
\label{Tunable-Raman}
\end{figure}

Tuning the laser photon energy, $E_{\rm L}$, reveals that the Raman frequency of this new feature in Fig.~\ref{Rayleigh-Raman}(a) is highly dispersive.  Figure \ref{Tunable-Raman}(a) shows the spectra obtained at various incident laser energies $E_{\rm L}$ as a function of the energy of the scattered photon $E_{\rm S}$. Although the Raman shift of the newly observed ERS feature changes with $E_{\rm L}$, the corresponding $E_{\rm S}$, remains centered around $2.08~\rm eV$ for $E_{\rm L}$ in the range 2.10~eV to 2.20~eV, and around 2.19~eV for $E_{\rm L}=2.33~{\rm eV}$. These values of $E_{\rm S}$ match the $M_{22}^-$ and $M_{22}^+$ energies obtained from the Rayleigh spectrum, respectively.  For a collection of more than 15 structure-assigned $M{\rm -SWNTs}$ excited slightly above a given $M_{ii}$ resonance, we have consistently observed a prominent feature, peaked at $M_{ii}$, in the Raman spectra.


The ERS feature has also been observed for several $M{\rm -SWNTs}$ lying on a substrate and for $M{\rm -SWNTs}$ that are part of a small bundle. In these cases, the $M_{11}$ or $M_{22}$  transitions have been determined from the RBM resonance window ~\cite{sonPRB2006}.  Again, we systematically observe a broad ERS feature at the energy of the $M_{ii}$ transitions, when $E_{\rm L}$ is near $M_{ii}$.  Interestingly, the ERS feature is also present in the anti-Stokes spectrum when a $M{\rm -SWNT}$ is excited slightly below resonance (Fig.~\ref{Tunable-Raman}(c)).   On the other hand, the Raman background of more than 15 structure-assigned $S{\rm -SWNT}$ excited near their $S_{ii}$ transitions is nearly flat, with the exception of a weak tail near the Rayleigh peak, as shown in the inset of Fig.~\ref{Rayleigh-Raman}(b).  Based on this survey of a statistically significant number of individual tubes, we conclude that the ERS feature is a hallmark of $M{\rm -SWNTs}$ and can be observed for any $M{\rm -SWNT}$ excited near resonance.

We attribute the newly observed peaks to a \textit{resonant} electronic Raman scattering process, involving low-energy electronic transitions across the linear subbands of $M{\rm -SWNTs}$. Figure~\ref{Tunable-Raman}(d) shows a schematic diagram of the resonant ERS process.  A continuous range of available electronic excitations may scatter light.  For scattering a photon by a first-order Raman process, only vertical electron-hole ({\rm e-h}) excitations ($q_{\rm e-h}=0$) satisfy the conservation of momentum.  Since the Coulomb interaction is long range, its contribution peaks sharply near $q_{\rm e-h}=0$ \,\cite{jiangPRB2007}, and therefore it is most significant for these vertical {\rm e-h} excitations. While the density of states for these excitations is constant~\cite{saitoBOOK}, the ERS events that result in outgoing photons of the same energy as one of the $M_{ii}$ transitions will be resonantly enhanced and will dominate the signal.  Consequently, the ERS feature is always centered at $M_{ii}$, irrespective of $E_{\rm L}$.  
  
When exciting a $M{\rm -SWNT}$ very close to an $M_{ii}^{}$ optical transition (Fig.~\ref{Tunable-Raman}), ERS processes, involving {\rm e-h} pairs with near-zero energy, are resonantly enhanced. Thus, a strong but featureless ERS tail is observed near the laser line and overlaps with the RBM feature ($E_{\rm L}=2.07~{\rm eV}$ for the (23,14) ${\rm SWNT}$  or $E_{\rm L}=2.00~{\rm eV}$ for (29,5)).  As $E_{\rm L}$ is tuned further away from $M_{ii}^{}$, higher energy {\rm e-h} pairs will contribute to the Raman spectrum and a well-defined ERS peak will develop at a Raman shift corresponding to the laser detuning ($E_{\rm L}=2.14~{\rm eV}$ for (23,14) or $E_{\rm L}=2.12~{\rm eV}$ for a (29,5) ${\rm -SWNT}$).  The energy of the scattered photons corresponding to the ERS peak, however, remains unchanged.

In this picture, the ERS spectrum is composed of contributions from many {\rm e-h} excitations, whose relative amplitudes are modulated by the nearest $M_{ii}$ resonance. In agreement with this, we find that the widths of the observed ERS features are of the same order of magnitude, as the associated $M_{ii}$ transitions.

The fact that the new ERS features are observed exclusively at $M_{ii}$ in $M{\rm -SWNTs}$, and not in $S{\rm -SWNTs}$, together with the presence of the ERS features both on the Stokes- and anti-Stokes sides of the spectra, rules out fluorescence as a possible interpretation.  Stokes-fluorescence from $M{\rm -SWNTs}$ or from higher-order transitions $(S_{ii},i>1)$ in $S{\rm -SWNTs}$ is extremely inefficient because the exciton radiative decay rate in SWNTs ($\approx0.1~\rm ns^{-1}$~\cite{perebeinosNL2005,spataruPRL2005,berciaudPRL2008}) is at least five orders of magnitude slower than non-radiative interband relaxation processes~\cite{manzoniPRL2005, lauretPRL2003,hertelAPA2002}. Anti-Stokes fluorescence would require additional phonon-assisted processes, which are even less probable. Also, the laser power was kept sufficiently low to ensure that the spectra were collected in the low excitation regime and therefore thermally induced effects can also be excluded.  Finally, in a fluorescence scenario, we expect to observe visible emission from both $M{\rm -SWNTs}$ \textit{and} $S{\rm -SWNTs}$, obviously contradicting our results.

\begin{figure}
		\includegraphics[width=0.45\textwidth]{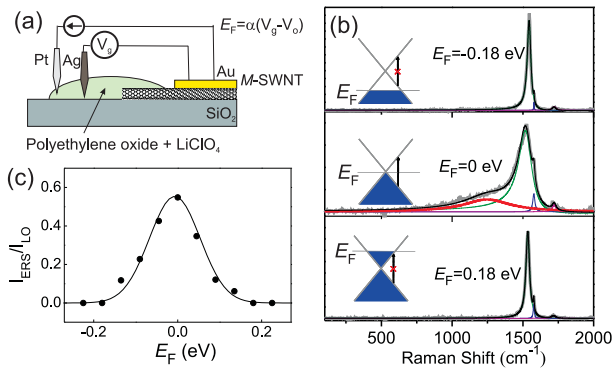}
		\caption{		
(Color online) (a)  A schematic of the electrochemical gating setup.  A Pt counter-electrode is used to maintain the gate voltage $V_{\rm g}$ between the Au contacted $M{\rm -SWNT}$ and a Ag pseudo-reference electrode.  The gate efficiency $\alpha$, and charge neutrality voltage $V_o$ relate $V_{\rm g}$ to $E_{\rm F}$~\cite{supp-info}. (b) The Raman spectra of an electrochemically gated $M{\rm -SWNT}$. The ERS peak shown by the bold red line is suppressed when the Fermi energy, $E_{\rm F}$, is shifted to positive and negative values.  The phonon feature is fit to a Fano lineshape for the lower-energy longitudinal optical (LO) mode and a Lorentzian for the narrower, higher-energy, transverse optical (TO) mode~\cite{wuPRL2007}.  All spectra are normalized to the LO mode intensity. (c) The ratio of the integrated intensities of the ERS and LO phonon peaks as a function of $E_{\rm F}$.  The solid curve is a Gaussian fit to guide the eye.}
\label{ERSgate}
\end{figure}

If low-energy electronic excitations are indeed responsible for the observed ERS features, then we should be able to suppress these features by changing the occupancy of the relevant electronic states.  Figure~\ref{ERSgate} shows the evolution of the Raman spectrum of a $M{\rm -SWNT}$ on an SiO$_2$ substrate whose Fermi level, $E_{\rm F}$, is tuned by means of an electrochemical gate (Fig.~\ref{ERSgate}(a)) ~\cite{farhatPRL2007,supp-info}.  The ERS feature, located at $\sim$1200~${\rm cm^{-1}}$, has less intensity when $E_{\rm F}$ is shifted to positive or negative values (Fig.~\ref{ERSgate}(b)). This sensitivity arises from Pauli-blocking the possibility of generating the electron-hole pairs~\cite{lazzeriPRB2006,farhatPRL2007,wuPRL2007}. More quantitatively, the FWHM of the curve in Fig.~\ref{ERSgate}(c) ($\sim 150~{\rm meV}$ or ($\sim$1200~${\rm cm^{-1}}$) exactly matches the Raman shift of the ERS feature, \textit{i.e.}, the average energy of the {\rm e-h} pairs, that contribute to the resonant ERS process. Thus, the results of Fig.~\ref{ERSgate} further demonstrate that the {\rm e-h} excitations are involved in the ERS process, and also that relatively low levels of doping $(<0.005~e/{\rm C~atom})$ will suppress the ERS features. 

These same {\rm e-h} pairs have recently been shown to couple strongly to the zone-center longitudinal optical (LO) phonons of $M{\rm -SWNTs}$ resulting in phonon softening and lifetime broadening~\cite{lazzeriPRB2006,farhatPRL2007,wuPRL2007}.  Given this coupling, the coherent addition of the ERS and LO phonon spectra could give rise to a Fano interference, thus accounting for the widely reported, albeit weak, asymmetry of the LO phonon feature in the Raman spectrum of $M{\rm -SWNTs}$ \,\cite{BrownPRB2001,oroncarlNL2005,pailletPRL2005,wuPRL2007}.  An analysis of this asymmetry vs.  its spectral overlap with the ERS feature is presented elsewhere.
  
In conclusion, we have shown that resonant electronic Raman scattering features can be observed from any  $M{\rm -SWNTs}$ with a standard micro-Raman spectrometer. The ERS features reveal a wealth of information about the 1D electronic states of $M{\rm -SWNTs}$.  First, due to the resonant nature of ERS, the spectral position of an ERS feature provides a direct measurement of the nearest $M_{ii}$ optical transitions. Thus, ERS can be used for a more accurate ``Raman-based'' structural assignment of individual $M{\rm -SWNTs}$. As an example, the $M{\rm -SWNT}$ in Fig.~\ref{Tunable-Raman}(c) is assigned to a (15,6) nanotube based on both the location of the ERS peak energies and its RBM frequency.  


More importantly, the ERS spectrum may also yield information about the continuum of low-energy electronic excitations by which the light is scattered.  By using a sharper laser cut-off, the lower energy side of the ERS could potentially be used to study the electronic bands of nominally metallic carbon nanotubes in the very low energy range, where minigaps due to curvature and electron correlation effects have been reported \,\cite{deshpandeSCIENCE2009}. We do not expect contributions to the ERS spectrum for energies within these minigaps.



The authors  thank V.V. Deshpande, R. Caldwell and J. Hone for help with sample preparation.  Work was carried in the Spectroscopy Laboratory supported by Grant NSF-CHE 0111370 and Grant NIH-RR02594.  H.F. and J.K. acknowledge the Center for Excitonics, funded  by DE-SC0001088.  M.S.D. and H.F acknowledge support from NSF/DMR-10-04197.  R.S. acknowledges MEXT grant No. 20241023.  M.K. acknowledges support from the Czech Ministry of Education,Youth and Sports (contract No. ME09060).  S.B. and T. F. H. acknowledge support from the NSF under grant CHE-06-41523, from NYSTAR, and from the Keck Foundation.


\end{document}